\documentclass[aps,english,10pt,a4,superscriptaddress,twocolumn]{revtex4}  
\usepackage{amsmath}
\usepackage{amsthm}
\usepackage{amssymb}
\usepackage{amsfonts}
\usepackage{bbm}
\usepackage{epsfig}
\usepackage{times}
\usepackage{babel}
\usepackage{color}
\usepackage{hyperref}
\usepackage{framed}
\usepackage{changes}
\usepackage{float}

\def\tr{{\rm tr}}

\def\R{{\mathbb{R}}}

\renewcommand{\vec}[1]{\mathbf{#1}}

\begin{document}
\title{Generalized eigenstate typicality in translation-invariant quasifree fermionic models}
\author{Jonathon Riddell}
\affiliation{Department of Applied Mathematics, University of Western Ontario, London, ON N6A 5BY, Canada}
\author{Markus P.\ M\"uller}
\affiliation{Institute for Quantum Optics and Quantum Information, Austrian Academy of Sciences, Boltzmanngasse 3, A-1090 Vienna, Austria}
\affiliation{Department of Applied Mathematics, University of Western Ontario, London, ON N6A 5BY, Canada}
\affiliation{Department of Philosophy, University of Western Ontario, London, ON N6A 5BY, Canada}
\affiliation{Perimeter Institute for Theoretical Physics, Waterloo, ON N2L 2Y5, Canada}

\date{January 16, 2018}

\begin{abstract}
We demonstrate a generalized notion of eigenstate thermalization for translation-invariant quasifree fermionic models: the vast majority of eigenstates satisfying a finite number of suitable constraints (e.g.\ fixed energy and particle number) have the property that their reduced density matrix on small subsystems approximates the corresponding generalized Gibbs ensemble. To this end, we generalize analytic results by Lai and Yang (Phys.\ Rev.\ B \textbf{91}, 081110 (2015)) and illustrate the claim numerically by example of the Jordan-Wigner transform of the XX spin chain.
\end{abstract}

\maketitle

\section{Introduction}
The old question of how closed quantum systems thermalize has recently experienced a resurgence of interest, motivated by novel experiments with ultracold atomic gases~\cite{BlochZwerger} as well as by new analytical insights from quantum information theory~\cite{GogolinEisert,PopescuShortWinter,Goldstein,Reimann,Linden,GemmerMichelMahler}. One major conjecture that is supposed to yield central insights into this problem is the Eigenstate Thermalization Hypothesis (ETH)~\cite{Deutsch,Srednicki,Rigol,RigolReview}. While there are different versions of the ETH, we focus on one formulation that has been considered, for example, in Refs.~\cite{Biroli,Mori,Mueller,Nandy,Brandao}: namely, that energy eigenstates of quantum many-body systems have expectation values on local observables (of small subsystems) that agree with  those of the canonical ensemble at the corresponding temperature (with a deviation that ideally goes to zero in the thermodynamic limit). Or, in a nutshell, \emph{eigenstates are locally thermal}. While some systems, in particular many-body localized systems~\cite{Nandkishore}, are known to violate the ETH, the hypothesis is expected to hold in different versions under certain natural regularity assumptions including translation-invariance and non-integrability~\cite{Shiraishi}.

Regardless of the specific formulation of the ETH that one is interested in, there are two cases that need to be distinguished. First, some models are known to satisfy a ``strong'' version of the ETH, in the sense that \emph{all} eigenstates are locally thermal. This behavior has been shown numerically for some non-integrable models~\cite{Rigol,Mondaini,Kim},  but there is currently no known analytic proof of this hypothesis which would uniformly apply to a large class of such models. However, significant advances have been made in understanding the physical implications of the strong ETH, and in developing numerical methods to test it~\cite{Steinigeweg,Anza,DePalma,Khodja,Sorg}.

Second, there is another important class of systems which possesses a (possibly large) number of conserved quantities in addition to the energy, in particular local or extensive quantities. For such integrable systems, the ETH cannot always hold in the strong sense: if two eigenstates of comparable energies differ in the values of some local conserved quantity, for example, they cannot both be locally close to the thermal state of the corresponding temperature. These models can still satisfy a weak version of the ETH, in the sense that the vast majority of eigenstates --- but not all --- are locally thermal. This weak version of the ETH has been rigorously proven for a large class of translation-invariant models, integrable or not~\cite{Biroli,Mori,Iyoda}, and for quasifree fermionic models~\cite{Lai,Vidmar} as well as the XXX spin chain~\cite{Alba}.

A more general approach for understanding eigenstate thermalization in integrable models has been to replace the canonical ensemble by a generalized Gibbs ensemble (GGE)~\cite{VidmarRigol,RDYO,RMO}. The GGE has first been employed in dynamical situations, where one is interested in understanding relaxation following a quantum quench~\cite{EisertFriesdorfGogolin}. It is defined as
\[
   \rho_{\rm GGE}=\frac 1 Z \exp\left(-\sum_i \beta_i \hat Q_i\right),
\]
where the $\hat Q_i$ denotes the set of relevant conserved quantities (which exist in particular due to integrability), and the $\beta_i$ are Lagrange multipliers that are chosen such that the expectation values $\langle \hat Q_i\rangle$ are equal to predefined initial values $Q_i$. If the initial state after a quantum quench is $|\psi\rangle$ then $Q_i=\langle  \psi|\hat Q_i|\psi\rangle$, and the number of $\hat Q_i$ in integrable models is typically very large (for example, it encompasses all mode occupation numbers in free models). The success of the GGE has motivated the formulation of a weak generalized version of the ETH (GETH)~\cite{Cassidy}: namely, that the vast majority of eigenstates with similar values of all relevant conserved quantities are locally close to the corresponding GGE. It has been claimed that the GETH fails in the case of the XXZ model~\cite{Pozsgay,Wouters}, but later work~\cite{IlievskiComplete} has shown that this was due to an incomplete choice of conserved quantities in the definition of the GGE: if additional quasilocal conserved quantities~\cite{Ilievski} are included, then the GETH holds as initially conjectured. This shows that the right choice of conserved quantities in the definition of the GGE can be a subtle issue.

In this paper, we aim to shed some light on the validity of the weak GETH by example of the analytically and numerically most accessible integrable models, namely translation-invariant quasifree fermionic models. We consider a version of eigenstate typicality that is in some sense ``in between'' the weak ETH and the weak GETH: eigenstates $|E\rangle$ are drawn at random according to fixed values of $n$ suitable conserved quantities $\hat Q_i,\ldots, \hat Q_n$, where $n$ is typically much smaller than the total number of relevant conserved quantities of the system (we consider $n$ to be constant and not to grow with system size; for example, $n=1$ corresponds to the weak ETH). Generalizing a result by Lai and Yang~\cite{Lai}, we prove analytically that the vast majority of eigenstates that satisfy those $n$ constraints is locally close to the corresponding GGE. In this sense, quasifree fermionic models satisfy generalized eigenstate typicality. We also illustrate our results numerically by example of the XX spin chain, which can be translated into a fermionic model by means of a Jordan-Wigner transformation.

Thus, translation-invariant quasifree fermionic models are concrete examples for which a version of the weak GETH can be analytically proven, and they represent interesting toy models for studying the impact of integrability on the different notions of eigenstate thermalization.

\section{Quasifree fermionic models}
\label{SecQuasifree}
In this paper, we consider quasifree fermionic models~\cite{EisertCramerPlenio} on a $d$-dimensional cubic lattice $\mathbb{Z}^d$. For simplicity we restrict ourselves to Hamiltonians of the form
\[
   \hat H= \sum_{j,k} h_{j,k} \hat f_j^\dagger \hat f_k
\]
on finite cubic regions, where the indices $j$ (and $k$) label the $L^d$ elements of the cube $\{1,\ldots,L\}^d$. That is, to every $1\leq j \leq L^d$ we associate a corresponding vector $\vec r_j\in\{1,L\}^d$ denoting the position of site $j$ (in one dimension, i.e.\ $d=1$, we have $\vec r_j=j$). We have the fermionic anticommutation relations
\[
   \{\hat f^\dagger_j,\hat f_k\}=\delta_{jk},\enspace \{\hat f^\dagger_j,\hat f^\dagger_k\}=\{\hat f_j,\hat f_k\}=0,
\]
and for $\hat H$ to be Hermitian we must have $h_{kj}=\bar h_{jk}$. We assume both translation-invariance and periodic boundary conditions, which can be expressed as
\[
   h_{jk}=h_{j'k'}\quad\mbox{if}\quad(\vec r_j-\vec r_k){\rm mod}\, L =(\vec r_{j'}-\vec r_{k'}){\rm mod}\, L,
\]
where the equation on the right-hand side is to be understood componentwise. It is well-known~\cite{VidmarRigol,Parkinson}, and can be checked by a straightforward calculation, that $\hat H$ can be diagonalized by introducing $L^d$ discrete momenta $\vec k\in\R^d$, where every $k_i$ is of the form $2\pi m_i/L$, with $1\leq m_i\leq L$ some integer. Again, we label these momenta by some integer $1\leq j \leq L^d$, such that the $j$-th momentum vector is $\vec k_j = 2 \pi \vec{r}_j/L$. The diagonalization is achieved by means of a discrete Fourier transform: defining
\[
   \hat d_{\vec k}:=\frac 1 {L^{d/2}} \sum_j e^{i\vec k \cdot \vec r_j}\hat f_j,
\]
the resulting operators $\hat d_{\vec k}$ in turn satisfy the fermionic anticommutation relations. They  allow us to rewrite the Hamiltonian in the form
\[
   \hat H=\sum_{\vec k} \epsilon_{\vec k} \hat d_{\vec k}^\dagger \hat d_{\vec k},
\]
where the $\epsilon_{\vec{k}}$ are (real-valued) energies of momentum excitations. Starting with the vacuum state which is identical in both the direct and reciprocal lattice, we can build all of the eigenstates using the fermionic momentum operators $\hat{d}_{\vec{k}}^\dagger$. In this paper we will focus only on those eigenstates that can be obtained directly by applying the creation operators, and we will not consider eigenstates that are superpositions of such states which exist if $\hat H$ is degenerate (as, for example, in the case of the XX model discussed in Section~\ref{SecXX}).

\section{Generalized eigenstate typicality}
\label{SecGET}
We now turn to a description in terms of Majorana operators. These are defined in terms of the fermionic creation and annihilation operators as
\begin{eqnarray*}
	\check{f}_{j,1} &=& \frac{1}{\sqrt{2}}(\hat{f}_j^{\dagger}+\hat{f}_j), \\
	 \check{f}_{j,2} &=& \frac{-i}{\sqrt{2}}(\hat{f}_j^{\dagger}-\hat{f}_j).
\end{eqnarray*}
They satisfy the anticommutation relation $\{\check{f}_{j,a},\check{f}_{k,b}\} = \delta_{jk}\delta_{ab}$. Analogously we define Majorana operators for momentum space as $\check{d}_{\vec{k},1}$ and $\check{d}_{\vec{k},2}$. The eigenstates of free fermionic models are Gaussian, thus describing the covariances of these operators is sufficient to completely define any eigenstate $\rho$ of the system. The covariance matrices with respect to the Majorana operators $\check f_{j,a}$ are defined as
\[
   \Gamma_{(j,a),(l,b)}^f =i \langle [ \check{f}_{j,a}, \check{f}_{l,b} ] \rangle
   = 2 i\langle \check f_{j,a}\check f_{l,b}\rangle - i \delta_{jl}\delta_{ab},
\]
where the expectation value is taken with respect to the state $\rho$, and $a,b\in\{1,2\}$. Analogously we define the covariance matrix $\Gamma_{(\vec k,a),(\vec k',b)}^d$ with respect to the $\check d_{\vec k,a}$.

Since we know the form of the eigenstates in momentum space, we begin with the reciprocal space covariances and investigate the local covariances through a suitable transformation. As we can construct all eigenstates through momentum creation operators $\hat{d}_{\vec{k}}$, the only non-zero entries of the covariance matrix of any eigenstate are (see also~\cite{Wilming})
\begin{equation}\label{exp1}
   \Gamma^d_{(\vec k,2),(\vec k,1)} = 2i\langle \check d_{\vec k,2} \check d_{\vec k,1}\rangle = -\Gamma^d_{(\vec k,1),(\vec k,2)}.
\end{equation}
For the purpose of investigating eigenstate thermalization, we partition our system with lattice sites $\{1,\ldots,L\}^d$ into two subsystems $A$ and $B$ such that $A \ll B$. The goal is to see if the eigenstates of our free fermionic model are locally close to thermal on the subsystem $A$ under suitable constraints. We thus investigate the covariances $\Gamma^f_{(j,a),(l,b)}$ for $j,l\in A$. The relation between the direct and reciprocal lattice will be given by a linear map $W$ which is found by applying the Fourier transform to the fermion operators and tracking its effects on the Majorana operators. Therefore
\[
   \check d_{\vec k,a} = \sum_{j,b} W_{(\vec k,a),(j,b)} \check f_{j,b},
\]
and we get
\begin{eqnarray*}
   W_{(\vec k,1),(j,1)} &=& \frac{1}{\sqrt{L^d}} \cos(\vec r_j\cdot \vec k)=W_{(\vec k,2),(j,2)},\\
   W_{(\vec k,1),(j,2)} &=& \frac{1}{\sqrt{L^d}} \sin(\vec r_j\cdot \vec k) = -W_{(\vec k,2),(j,1)}.
\end{eqnarray*}
This transformation inherits orthonormality from the Fourier transform. Using this transformation, we express the local covariance $\Gamma^f_{(j,2),(l,1)}$ in terms of momentum space covariances,
\[
   \Gamma^f_{(j,a),(l,b)} = \sum_{\vec k,c} W_{(\vec k,c),(j,a)} \sum_{\vec k',d} \Gamma^d_{(\vec k,c),(\vec k',d)} W_{(\vec k',d),(l,b)}.
\]
There are four different cases of $a,b\in\{1,2\}$; we will now consider the case $a=2$ and $b=1$, and give the results for the other cases at the end of this section (the derivations are analogous). Substituting~(\ref{exp1}) into this equation, we get
\begin{eqnarray*}
\Gamma^f_{(j,2),(l,1)}=2 i \sum_{\vec k}\langle \check d_{\vec k,2}\check d_{\vec k,1}\rangle\left(W_{(\vec k,2),(j,2)}W_{(\vec k,1),(l,1)} \right. \\
\left. - W_{(\vec k,1),(j,2)}W_{(\vec k,2),(l,1)}\right).
\end{eqnarray*}
Expanding and simplifying the trigonometric terms from $W$, and using $\hat{d}_{\vec k}^\dagger \hat{d}_{\vec k} - \frac{1}{2} = i \check{d}_{\vec k,2}\check{d}_{\vec k,1}$, we find
\begin{equation} \label{fgam}
\Gamma_{(j,2),(l,1)}^f	=  \frac{2}{L^d}\sum_{\vec k}\left(\langle \hat{d}_{\textbf{k}}^\dagger \hat{d}_{\textbf{k}} \rangle - \frac{1}{2}\right) \cos(\textbf{k}\cdot (\textbf{r}_j-\textbf{r}_l)).
\end{equation}
We can now proceed similarly to~\cite{Lai} and investigate the covariances of typical eigenstates. However, we will work in a more general setting, by allowing the eigenstates to be drawn at random according to several constraints, which modifies and generalizes the notion of typicality. As $L \to \infty$ we see that the values the vector components of $\textbf{k}$ form a dense set on $(0,2\pi]^d$. We  partition momentum space into a large number of cubes with side length $\Delta k=2\pi\ell/L$, where $1\ll \ell \ll L$. Since we take $L$ to infinity, it is sufficient to consider those $L$ such that $L=\ell\cdot s$, where $\ell,s$ are both integers. Each of these cubes contains $g=\ell^d\gg 1$ momentum points. In the thermodynamic limit, we demand that $g$ tends to infinity, but does so only sublinearly in $L$, i.e.\ $o(L)=g=\ell^d\to\infty$ as $L\to\infty$. In this way, we also enforce that for large $L$,
\[
   \frac{1}{L} \ll \Delta k \ll \frac{1}{L_A},
\]
where $L_A$ is the maximum displacement on an axis we might observe inside the subregion $A$ (we do not increase $A$ with $L$). Let us label the cubes by $C_m$, where $m\in\{1,\ldots,s^d\}$. This condition allows us to approximate
\[
   \cos(\textbf{k}\cdot (\textbf{r}_j-\textbf{r}_l)) \approx \cos(\textbf{k}_m\cdot (\textbf{r}_j-\textbf{r}_l)), \quad (\textbf{k} \in C_m),
\]
where $\vec{k}_m$ is an arbitrarily picked but fixed momentum in cube $C_m$. This is due to the enforced relation $\Delta k L_A \ll 1$. It is here where the condition $j,l\in A$ enters, i.e.\ the fact that we are looking at the reduced state on the subsystem $A$ only. With the approximations made we can simplify equation~(\ref{fgam}) by summing over the cubes,
\begin{equation} \label{agamma}
\Gamma_{(j,2),(l,1)}^f	\approx  \frac{2g}{L^d}\sum_m\left(n_m - \frac{1}{2}\right) \cos(\vec{k}_m\cdot (\vec{r}_j-\vec{r}_l))
\end{equation}
for all $j,l\in A$, where $n_m=\frac 1 g\sum_{\vec{k}\in C_m} \langle \hat d_{\vec k}^\dagger \hat d_{\vec k}\rangle$ is the density of momentum excitations in cube $C_m$. In fact, an elementary calculation shows that the absolute difference between~(\ref{agamma}) and~(\ref{fgam}) is upper-bounded by $\mathcal{O}(\ell L_A/L)$. Thus, in order for our approximations to be valid, $L_A$ must grow less than linearly with $L$, confirming the results of~\cite{Lai}. Since we keep $A$ fixed and do not change it with $L$, this is satisfied in our case.

The values that $n_m$ can take in the thermodynamic limit will become dense in the unit interval so that we will later be able to take a derivative with respect to $n_m$. With this notation it is easy to see why several microstates of different momentum excitation arrangements will look locally identical, as different distributions of $\langle \hat{d}^\dagger_{\textbf{k}} \hat{d}_{\textbf{k}} \rangle$ can lead to the same distribution of $n_m$. We now collect a finite set of conserved quantities $\hat Q_1,\ldots,\hat Q_n$ linear in momentum space number operators which commute with the Hamiltonian,
\begin{equation}
 \hat{Q}_i = \sum_{\textbf{k}}q_{i,\textbf{k}}\hat{d}^\dagger_{\textbf{k}} \hat{d}_{\textbf{k}}.
 \label{eqConstraints}
\end{equation}
In the following, we need that the $q_{i,\vec{k}}$ do not vary too wildly in $\vec{k}$. Therefore, we impose the condition that these coefficients are uniformly bounded, i.e.\ there is some constant $C$ such that $|q_{i,\vec{k}}|\leq C$ for all $\vec{k}$. Furthermore, we assume that these coefficients are Lipschitz continuous in $\vec{k}$, except possibly within $\mathcal{O}(L^{d-1})$ many cubes $C_m$ (recall that the total number of these cubes is $\mathcal{O}(L^d)$). Here this means that there is some constant $c$ such that
\begin{equation}
   |q_{i,\vec{k}}-q_{i,\vec{k}'}|\leq c\cdot \Delta k\quad\mbox{for all }\vec{k},\vec{k}'\in C_m
   \label{eqApproxQ}
\end{equation}
for all but $\mathcal{O}(L^{d-1})$ many cubes $C_m$. While we allow that the $q_{i,\vec{k}}$ depend on $L$, we will only consider examples where $q_{i,\vec{k}}$ is either constant in $L$ or converges to some fixed function in the limit $L\to\infty$. Therefore, we may and will assume that the constants $c$ and $C$ are independent of $L$. Since the number of different $\vec{k}$ grows like $L^d$, this means that the $\hat Q_i$ describe extensive quantities. For example, for $q_{i,\vec{k}}=1$, we recover the total particle number, and for $q_{i,\vec{k}}=\epsilon_{\vec k}$ we recover the total energy. Furthermore, this implies that we can approximate the values of $q_{i,\vec{k}}$ inside the cubes $C_m$ in the following way. For every $m$, let $\vec{k}_m\in C_m$ be the arbitrarily chosen momenta from further above, and set $q_{i,m}:=q_{i,\vec{k}_m}$. Thus, for all but $\mathcal{O}(L^{d-1})$ many cubes $C_m$, we have
\[
   q_{i,\vec{k}}\approx q_{i,m}\quad\mbox{for all }\vec{k}\in C_m,
\]
and the difference is bounded as in~(\ref{eqApproxQ}).

Let us now fix some values $Q_i$ (which are real numbers), and consider the set of all those eigenstates with momentum excitation densities $n_m$ such that
\begin{equation}
    Q_i = g\sum_{m}n_m q_{i,m}.
    \label{eqFix}
\end{equation}
Since $\langle \hat Q_i\rangle$ is approximately equal to the right-hand side of this equation, this will pick out eigenstates that have  approximately fixed expectation values $\langle \hat Q_i\rangle\approx Q_i$. In more detail, one easily verifies that the conditions above imply
\begin{equation}
   \frac 1 {L^d} \langle \hat Q_i\rangle -\frac 1 {L^d} Q_i \stackrel{L\to\infty}\longrightarrow 0,
   \label{eqSameDensities}
\end{equation}
i.e.\ the densities of these extensive quantities converge to each other in the thermodynamic limit. We are interested in \emph{typical} eigenstates that satisfy~(\ref{eqFix}), and thus consider all those eigenstates as equally probable. Thus, in each cube, we distribute $g\cdot n_m$ excitations uniformly over all the $g$ possible modes. The number of available microstates becomes
\[
 W_m = {g \choose {g n_m}}=\frac{g!}{(gn_m)!(g-gn_m)!},
\]
and the total number of accessible microstates is $W=\prod_m W_m$. The distribution for $n_m$ we are interested in is the most probable one, that is, the one at the peak of the distribution of $n_m$. Thus we introduce Lagrange multipliers $\lambda_i$ and maximize $W$ under the constraints~(\ref{eqFix}),
\[
   \frac{\partial}{\partial n_m}\left(\ln W -\sum_{i} \lambda_i\, g\sum_{m}q_{i,m}n_m \right)=0.
\]
A straightforward computation, using Stirling's approximation (valid since $g\gg 1$), yields
\begin{equation} \label{fermidirac}
 n_m =  \frac{1}{1+e^{\sum_{i}\lambda_iq_{i,m}}}.
\end{equation}
Thus we observe a distribution which, generalizing~\cite{Lai}, resembles a generalized Fermic-Dirac distribution. If we substitute eq.~(\ref{fermidirac}) into eq.~(\ref{agamma}) and use the fact that the $q_{i,\vec{k}}$ are piecewise continuous, the sum tends to an integral in the thermodynamic limit, namely
\begin{equation} \label{intgamma}
\small
\Gamma_{(j,2),(l,1)}^f = \frac{2}{(2 \pi)^d} \int\left( \frac{1}{1+e^{\sum_{i}\lambda_iq_{i,\vec k}}}-\frac{1}{2}\right)\cos( \vec{k} \cdot  (\vec{r}_l-\vec{r}_j))d\vec{k}.
\end{equation} 
We will now compare this to the covariance matrix of a generalized Gibbs ensemble (GGE). The correct state to compare to is not the local GGE, but the reduction of the global GGE. This has been observed in several recent works, where either global eigenstates or global microcanonical states have been shown to be locally close to the \emph{local reduction of the global thermal state}~\cite{Riera,Mueller}, not necessarily to the local thermal state (witnessed also by the use of \emph{intensive local} observables instead of local observables~\cite{Biroli}). The global GGE covariance matrix has arbitrary conserved quantities expressed  in terms of Majorana operators. In this new context the conserved quantities need to be rephrased in terms of Majorana operators.
\begin{eqnarray*}
\hat{Q}_i &=& \sum_{k} q_{i,\vec k} \hat{d}_{\vec k}^\dagger\hat{d}_{\vec k} \\
&=& \frac{i}{2}\sum_{k}q_{i,\vec k}(\check{d}_{\vec k, 2}\check{d}_{\vec k,1}-\check{d}_{\vec k, 1}\check{d}_{\vec k,2}) +\frac{1}{2}\sum_{k}q_{i,\vec k}.
\end{eqnarray*}
Let us throw away the state-independent offset $\frac{1}{2}\sum_{k}q_{i,\vec k}$, making our new set of conserved quantities
\[
\hat{Q}_i -\frac{1}{2}\sum_{k}q_{i,\vec k} = \hat{Q}_i'.
\]
This adjustment to the observables plays an analogous role to the $-\frac{1}{2}$ term in~(\ref{intgamma}). Then the GGE is defined by the density matrix
\[
	\rho_{\textsc{GGE}} = \frac{e^{-\sum_{i}\beta_i \hat Q_i'}}{Z},
\]
where $Z=\tr(\exp(-\sum_i\beta_i \hat Q'_i))$ and the $\beta_i$ are chosen such that $\tr(\rho_{\rm GGE}\hat Q_i)=Q_i$. This gives us the covariances
\[
\small
\Lambda_{(\vec k,a),(\vec k',b)}^d = \left\{
\begin{array}{cl}
   \tanh(\frac{1}{2}\sum_{i}\beta_i q_{i,\textbf{k}} ) & (a,b)=(1,2), \vec{k}=\vec{k'},  \\
 - \tanh(\frac{1}{2}\sum_{i}\beta_i q_{i,\textbf{k}} ) &(a,b)=(2,1), \vec{k}=\vec{k'},\\
 0 &\text{otherwise}.
\end{array}\right.
\]
The expression for $\Lambda_{(j,2),(l,1)}^f$ is found with an analogous method as $\Gamma_{(j,2),(l,1)}^f$. When followed through we arrive at
\[
\Lambda_{(j,2),(l,1)}^f = -\frac{1}{L^d}\sum_{\textbf{k}} \tanh\left(\frac{1}{2}\sum_{i}\beta_i q_{i,\textbf{k}} \right) \cos( \textbf{k}\cdot(\textbf{r}_l-\textbf{r}_j)).
\]
In the thermodynamic limit we can again express this sum as an integral over momentum space:
\begin{equation} \label{thermGGE}
\small
\Lambda_{(j,2),(l,1)}^f = -\frac{1}{(2\pi)^d} \int\tanh\left(\frac{1}{2}\sum_{i}\beta_i q_{i,\textbf{k}} \right) \cos( \textbf{k}\cdot(\textbf{r}_l-\textbf{r}_j))d\textbf{k}.
\end{equation}
Noting that $\frac 1 {1+e^x} - \frac 1 2 =-\frac 1 2 \tanh \left(\frac 1 2 x\right)$, we see that~(\ref{thermGGE}) is identical to~(\ref{intgamma}) for $j,l\in A$, except that the $\lambda_i$ are replaced by $\beta_i$. While this treats the case $(a,b)=(2,1)$, the same method shows that we get identical forms also for the other values of $a,b$. Specifically,
\begin{widetext}
\begin{align*}
\Gamma^f_{(j,1),(l,2)}=-\Gamma^f_{(j,2),(l,1)}, \qquad
\Gamma^f_{(j,1),(l,1)} = \Gamma^f_{(j,2),(l,2)}=\frac{1}{(2\pi)^d} \int\tanh\left(\frac{1}{2}\sum_{i}\lambda_i q_{i,\vec k} \right) \sin( \vec{k}\cdot(\vec{r}_j-\vec{r}_l))d\textbf{k}.
\end{align*}
\end{widetext}
This proves that typical eigenstates under linear constraints of the form~(\ref{eqConstraints}) are locally (on small subsystems $A$) close to a generalized Gibbs ensemble of a suitable choice of temperature and other Lagrange multipliers, and in fact identical to it in the thermodynamic limit. We do not formally prove this, but we expect that $|\lambda_i-\beta_i|\to 0$ for $L\to\infty$. This is because temperature and the other Lagrange multipliers are usually functions of the densities of the conserved quantities. As~(\ref{eqSameDensities}) shows, our eigenstates will in the thermodynamic limit have the same densities as the GGE with Lagrange multipliers $\beta_i$. On the other hand, we expect that the GGE which resembles the energy eigenstate on the subsystem $A$ will lead to the same densities for $L\to\infty$, and thus attain the same values of the Lagrange multipliers.

In the special case of two conserved quantities $\hat Q_1=\hat H$ (energy) and $\hat Q_2=\hat N$ (particle number), we recover the result from~\cite{Lai}. However, our results cover more general cases, and we will provide numerical examples in Section~\ref{SecNumerics}.

\section{Application to the XX model}
\label{SecXX}
To illustrate the analytic results of this paper numerically, we now proceed by discussing the one-dimensional $(d=1)$ XX spin chain~\cite{BarrySimon,VidmarRigol} with open boundary conditions. The Hamiltonian for this model is given by
\[
	\hat{H} = J \sum_{i=1}^{L-1} \left(S_{i}^{X}S_{i+1}^{X} + S_{i}^{Y}S_{i+1}^{Y}\right) -\lambda \sum_{j=1}^L S_{j}^{Z},
\]
where $S_i^X,S_i^Y$ and $S_i^Z$ are the standard spin-$1/2$ operators at site $i$, $L$ is the number of sites, $J$ is the interaction coefficient, $\lambda$ is the strength of the magnetic field applied to the $z$-axis, and we set $\hbar=1$ for convenience. Through a Jordan-Wigner transformation~\cite{Coleman}
\begin{eqnarray*}
   S_i^+ &=& \prod_{j=1}^{i-1} \left(1-2 \hat f_j^\dagger \hat f_j\right)\hat f_i^\dagger,\quad S_i^- = \prod_{j=1}^{i-1} \left(1-2 \hat f_j^\dagger \hat f_j\right) \hat f_i,\\
      S_i^Z &=& \hat f_i^\dagger \hat f_i-\frac 1 2,
\end{eqnarray*}
where $S_i^{\pm} = (S_i^X\pm i S_i^Y)/2$, we can rewrite the Hamiltonian in terms of the fermionic creation and annihilation operators $\hat f_j,\hat f_j^\dagger$ as
\[
	\hat{H} = \frac{J}{2} \sum_{i=1}^{L-1}\left( \hat{f}_i^{\dagger}\hat{f}_{i+1} + \hat{f}_{i+1}^{\dagger}\hat{f}_i\right) -\lambda\sum_{j=1}^L\left(\hat{f}_{j}^{\dagger}\hat{f}_{j}-\frac{1}{2}\right).
\]
While we recover a quasifree fermionic model of the form discussed in Section~\ref{SecQuasifree}, there are two problems --- which, however, turn out not to spoil our calculation. First, the Jordan-Wigner transformation does not completely preserve locality: the $j$th creation and annihilation operators of the transformed Hamiltonian are built from all of the sites $1,2,\ldots,j$ of the original spin Hamiltonian. Yet, if we consider subregions $A=\{1,2,\ldots,m\}$ (as we will do in our numerical calculations), then these blocks of sites are preserved by the Jordan-Wigner transformation. In other words, statements about the first $m$ sites of the quasifree fermionic model will directly translate to statements about the first $m$ sites of the original spin Hamiltonian.

Second, the fermionic Hamiltonian that we obtain has open boundary conditions, not periodic boundary conditions as assumed in Section~\ref{SecGET}. Nevertheless, since there are no finite-temperature phase transitions in $d=1$ dimensions~\cite{Araki69,Araki75}, we expect that boundary terms will become irrelevant in the thermodynamic limit for all questions of thermalization, including generalized eigenstate typicality as discussed here. Therefore, we expect to fully recover the analytical results of Section~\ref{SecGET} for the XX model. In particular, we will see below that we can still diagonalize the Hamiltonian by similar methods as in Section~\ref{SecGET}.

Next, for simplicity, we get rid of the constant $\lambda/2$ term in the Hamiltonian, obtaining a slightly modified version
\[
	 \hat{H}' = \sum_{i,j}^L M_{i,j}\hat{f}_i^{\dagger}\hat{f}_j.
\]
The next step is to diagonalize $M$, a banded Hermitian matrix which only has non-zero entries on its diagonal and on its immediate off diagonal entries; that is, $ M_{i,i} = -\lambda$ and $ M_{i,j} =J/2 $ if $|i-j|=1$. Invoking the Gershgorin circle theorem~\cite[Thm.\ 7.2.1]{Golub}, the eigenvalues $\epsilon_k$ must all lie in the circle about $-\lambda$ with radius J. Thus $|\epsilon_k+\lambda|\leq|J|$. With this expression we can assume the form $(\epsilon_k+\lambda)/J=\cos(a_k)$. Indeed one finds that the eigenvectors of $M$ (labelled by $j=1,\ldots,L$) are
\[
	\vec{v}_j=\left(\sqrt{\frac{2}{L+1}}\sin\left(\frac{j\pi  k}{L+1}\right)\right)_{ k=1,\dots L}
\]
and the energies of the modes are
\[
	\epsilon_{\vec k} = J\cos\left(\frac{L \vec{k}}{2(L+1)}\right)-\lambda,
\]
where $\vec k=2\pi k/L$ and $k=1,\ldots,L$. This allows us to write our Hamiltonian as
\[
	\hat{H} = \sum_{\vec{k}}\epsilon_{\vec{k}}\hat{d}_{\vec k}^\dagger\hat{d}_{\vec{k}},
\]
where these new fermion operators are defined as
\begin{eqnarray*}
	\hat{d}_{\vec{k}}^\dagger &=&\sqrt{\frac{2}{L+1}} \sum_{j=1}^{n}\sin\left(\frac{j\pi k}{L+1}\right)\hat{f}_j^\dagger, \\
	\hat{d}_{\vec{k}} &=&\sqrt{\frac{2}{L+1}} \sum_{j=1}^{n}\sin\left(\frac{j\pi k}{L+1}\right)\hat{f}_j.
\end{eqnarray*}
These new fermion operators obey the usual anti-commutation relations. Similarly as in Section~\ref{SecGET}, we will now transform this into the language of Majorana operators. It allows us to express the Hamiltonian as
\[
	\hat{H}' =  \frac{i}{2}\sum_{\vec{k}}\epsilon_{\vec{k}}(\check{d}_{\vec{k},2}\check{d}_{\vec{k},1}-\check{d}_{\vec{k},1}\check{d}_{\vec{k},2}) +\frac{1}{2}\sum_{\vec{k}} \epsilon_{\vec{k}}.
\]
One last time we modify the spectrum of our Hamiltonian and remove the constant term,
\[
	\hat{H}'' =  \frac{i}{2}\sum_{\vec{k}}\epsilon_{\vec{k}}(\check{d}_{\vec{k},2}\check{d}_{\vec{k},1}-\check{d}_{\vec{k},1}\check{d}_{\vec{k},2}).
\]
This gives us the Hamiltonian we will work with for the numerical experiments in the next section. The linear map between the two different sets of Majorana operators within this model is now given by 
\begin{align*}
	W_{(\vec{k},1),(j,1)}  = \sqrt{\frac{2}{L+1}}\sin\left(\frac{j L \vec{k}}{2(L+1)}\right) = W_{(\vec{k},2),(j,2)}, \\ 
	W_{(\vec{k},2),(j,1)}  = 0 = W_{(\vec{k},1),(j,2)}.
\end{align*}
The covariances of the eigenstates can be expressed in matrix notation as~\cite{Wilming}
\[
\Gamma^d = \bigoplus_{j=1}^L (-1)^{k_j} \begin{pmatrix}
0 & 1 \\
-1 & 0
\end{pmatrix}
\]
with $k_j = 0$ resp.\ $1$ representing a mode being empty resp.\ excited. The basis is chosen such that the $2\times 2$ blocks represent the entries $\Gamma^d_{(\vec{k},a),(\vec{k},b)}$ for $a,b=1,2$ and fixed $\vec{k}$, and antisymmetry of the blocks corresponds to $\Gamma^d_{(\vec{k},a),(\vec{k},b)}=-\Gamma^d_{(\vec{k},b),(\vec{k},a)}$. As we show in Appendix~\ref{SubsecGGE}, the covariances of the generalized Gibbs ensemble can similarly be expressed in matrix form as
\[
 \Lambda_{\rm GGE}^d = \bigoplus_{\vec{k}} \tanh\left(\frac{1}{2}\sum_{i}\beta_i q_{i,\vec{k}} \right) 
\begin{pmatrix}
 0 & 1 \\
 -1 & 0
 \end{pmatrix}.
\]
Both of the forms given are covariances for the mode Majorana operators $\check{d}_{\vec k,a}$. Noting that $W=W^T$ we can transform to local space via
\[
\Gamma^f = W \Gamma^d W.
\]
Finally, the last useful relation we will need for the numerics is a way to calculate the expectation value of an operator with a covariance matrix. We proceed similarly to~\cite{Wilming}, but for general operators. Suppose we have an observable
\[
	Q =  \frac{i}{2}\sum_{\vec{k}}q_{\vec{k}}(\check{d}_{\vec{k},2}\check{d}_{\vec{k},1}-\check{d}_{\vec{k},1}\check{d}_{\vec{k},2}).
\]
We can rewrite it as
\[
	Q = i\sum_{\vec{j},\vec{l}} \sum_{a,b} q_{(\vec{j},a),(\vec{l},b)}\check{d}_{\vec{j},a}\check{d}_{\vec{l},b}
\]
and thus
\begin{eqnarray*}
	\langle Q \rangle &=& \tr(\rho Q) = \sum_{\vec{j},\vec{l}} \sum_{a,b}\tr(\rho\, i\,q_{(\vec{j},a),(\vec{l},b)}\check{d}_{\vec{j},a}\check{d}_{\vec{l},b} ) \\
	&=& \frac{1}{2}\sum_{\vec{j},\vec{l}} \sum_{a,b}q_{(\vec{j},a),(\vec{l},b)}\Gamma^d_{(\vec{j},a),(\vec{l},b)} = -\frac{1}{2}\tr(q\Gamma^d).
\end{eqnarray*}
This equation allows us to solve for the Lagrange multipliers in the generalized Gibbs ensembles, and hence to investigate generalized eigenstate typicality in the one-dimensional XX model.

\section{Numerical results}
\label{SecNumerics}
In this section, we numerically test the notion of generalized eigenstate typicality against eigenstates sampled in a variety of ways from the XX-model in open boundary conditions. For the following numerical investigations we fix $J=1$ and $\lambda = \frac{1}{2}$, which is in the critical regime of parameters such that the model is gapless in the thermodynamic limit. The choice of parameters is more or less arbitrary, since the results from Section~\ref{SecGET} are expected to hold for all choices of parameters as discussed in Section~\ref{SecXX}. Yet, our choice of parameters avoids the special case $|\lambda|=J$ which would be on the boundary between gapped and gapless phase, and it describes a case where the magnetic field strength and the interaction are of comparable size. We use the results of the previous section to construct the eigenstate covariance matrix, measure its expectation values for the conserved quantities and build the corresponding generalized Gibbs ensemble. We define our subsystem $A$ as the first two fermion sites, that is, $A=\{1,2\}$, which translates to us investigating the statistics of the first two spin sites in our XX-model. The remaining $(L-2)$ lattice sites will be called $B$. Numerically we are interested in convergence behavior as we increase the number of lattice sites in $B$ with fixed number of sites in $A$.  The upper left corner of the local covariance matrices contain all of the local statistics for $A$ in the form of a $4\times 4$ submatrix. We thus define the local difference between the generalized Gibbs ensemble and an eigenstate as

\[
D = \sqrt{\sum_{j,k\in A}\sum_{a,b\in\{0,1\}} \left(\Gamma_{(j,a),(k,b)}^f-\Lambda_{(j,a),(k,b)}^f\right)^2}.
\]
Note that the momentum vectors $\vec{k}$ are now simply real numbers, and we can label them by integers $k$, such that $\vec{k}=2\pi k/L$. We are thus replacing the labels $\vec{k}$ in the covariance matrices by labels $k$.

This value of $D$ represents the distance between the local reduction of the given energy eigenstate $\Gamma\equiv \Gamma(L)$ and the local reduction of the (generalized) Gibbbs ensemble $\Lambda\equiv\Lambda(L)$ \emph{of the finite chain of length $L$}. Alternatively, one is often interested in the local difference between the eigenstate $\Gamma(L)$ and the \emph{thermodynamic (generalized) Gibbs state}, defined as $\Lambda(\infty):=\lim_{L\to\infty}\Lambda(L)$ (this convergence is understood in the weak sense, cf.~\cite{Mueller}). ``Eigenstate thermalization'' can either refer to the claim that $\Gamma^f(L)\approx \Lambda^f(L)$ locally, or to the claim that $\Gamma^f(L)\approx \Lambda^f(\infty)$ locally. However, it turns out that for the chains lengths $L$ and the parameters that we are probing in this section, the finite value of $L$ plays almost no role for the numerical results, so that we are basically testing both of these statements. That is, in the regime that we are probing,
\begin{equation}
   \Lambda_{(j,a),(k,b)}^f\equiv \Lambda_{(j,a),(k,b)}^f(L)\approx \Lambda_{(j,a),(k,b)}^f(\infty)
   \label{eqAlmostThere}
\end{equation}
for $j,k\in A$ and $a,b\in\{0,1\}$. Thus, $D$ is an excellent approximation to the local difference between the energy eigenstate and the (generalized) Gibbs state in the thermodynamic limit. Numerically, this can be seen by observing that our Lagrange multipliers never exceed $\beta_i\approx 6.5$ (in the canonical case, $\beta\lesssim 2.2$). On the other hand, as Table~\ref{TableFiniteSize} for the canonical ensemble exemplarily demonstrates, the differences between the $\Lambda_{(j,a),(k,b)}^f(L)$ for the different values of $L$ are numerically only significant for much larger values of $\beta$, which is evidence that~(\ref{eqAlmostThere}) is an excellent approximation in our regime.
\begin{table}[htb]
\centering
\begin{tabular}{r|c|c|c|}
& $L=100$ & $L=200$ & $L=300$ \\
\hline
$\beta=1$ & 0.231542 & 0.231542 & 0.231542 \\
\hline
$\beta=2$ & 0.388376 & 0.388376 & 0.388376 \\
\hline
$\beta=6$ & 0.570586 & 0.570586 & 0.570586 \\
\hline
$\beta=50$ & 0.608513 & 0.608513 & 0.608513 \\
\hline
$\beta=100$ & 0.60896 & 0.608877 & 0.608877 \\
\hline
$\beta=400$ & 0.613204 & 0.608976 & 0.608865 \\
\hline
$\beta=1000$ & 0.613974 & 0.608975 & 0.607782 \\
\hline
\end{tabular}
\caption{Entry $\Lambda^f_{(1,2),(1,1)}(L)$ of the canonical ensemble covariance matrix. Only for very high inverse temperatures $\beta$ are there any finite-size effects for the chain lengths $L$ that we are considering.}
\label{TableFiniteSize}
\end{table}

\subsection{Canonical ensemble / Gibbs ensemble (GE)}
We start by a quite naive test of eigenstate thermalization in its most simple formulation. Naively, one might expect that ``most'' eigenstates are locally thermal, i.e.\ close to the corresponding Gibbs ensemble.

\begin{figure}[h]
	\centering
	\includegraphics[width=\linewidth]{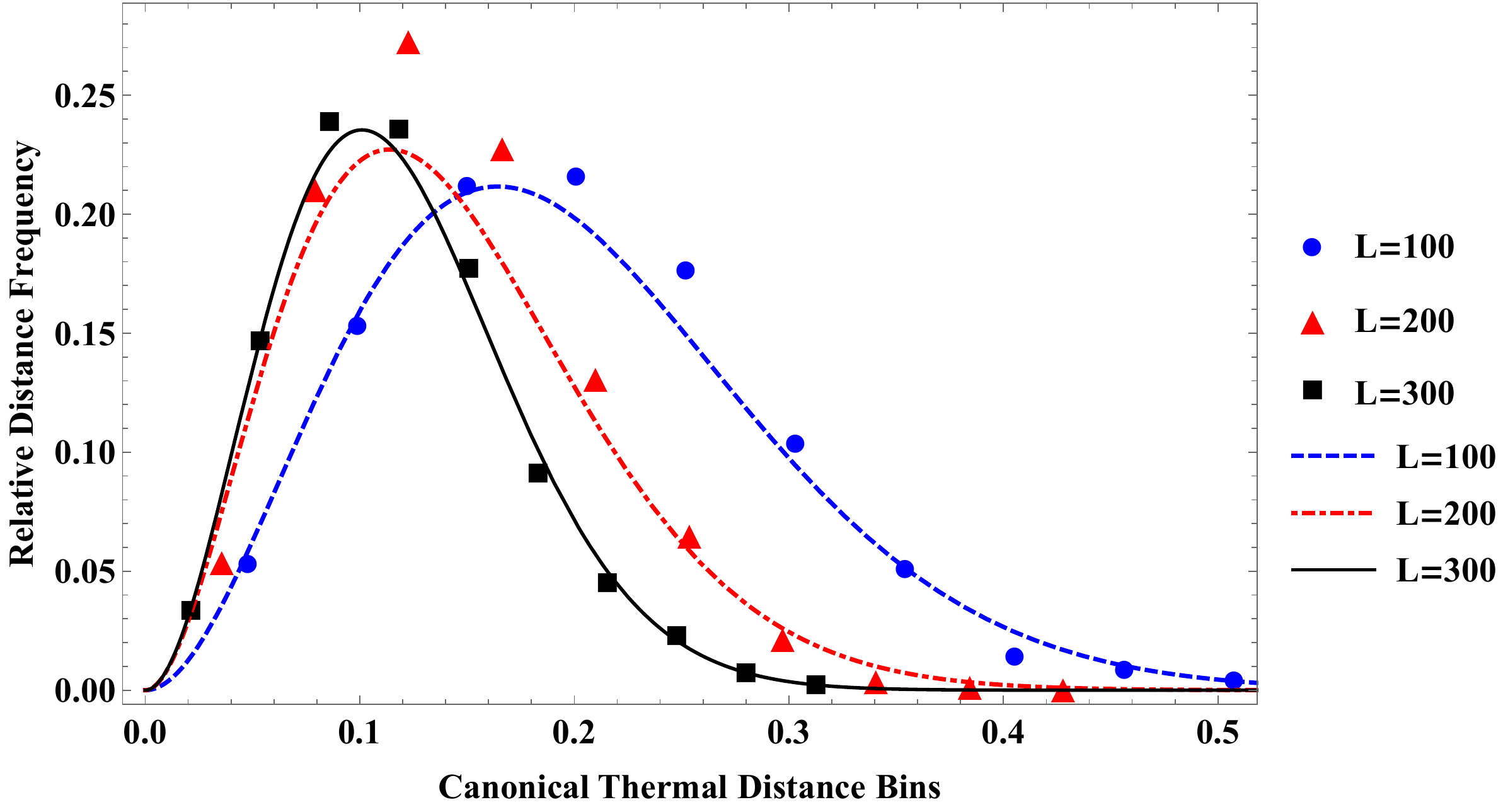}
	\caption{ Frequency plot of the local distance between a typically excited eigenstate and the Gibbs ensemble appearing within a given bin. The frequency plots in this section are constructed in the following way. 1800 eigenstates are sampled at a specific $L$. The local distance to the corresponding GGE is recorded. This data and is then sorted into 10 bins. The bins are constructed by taking the smallest  and largest distance observed and creating 10 equally spaced bins within the interval contained by these values. The frequency at which a distance is observed within these bins is then calculated from the data, and plotted on the vertical axis. The horizontal axis point is plotted as the mid point of the corresponding bin. The bin data and the frequency we observe is then fitted to a Maxwell-Boltzmann distribution of the form $f(x)=x^2 e^{\alpha + \beta x + \gamma x^2}$ with suitable $\alpha,\beta,\gamma\in\mathbb{R}$, which is  a natural distribution function that turns out to interpolate our data quite well.}
	\label{fig:GEhalf}
\end{figure}

Numerically, the easiest way to draw eigenstates at random is to generate them by applying random creation operators to the vacuum. Here we fix the \emph{excitation ratio} $r_{\rm exc}$ as a real number between $0$ and $1$. We then generate eigenstates with $n\cdot r_{\rm exc}$ excitations at random, determine their energy expectation value, and compare them locally (on $A$) to the corresponding Gibbs state of suitable inverse temperature $\beta$, with covariance matrix
\[
\Lambda_{GE}^d = \bigoplus_{k=1}^L \tanh\left(\frac{\beta\epsilon_k}{2}\right)
\begin{pmatrix}
0 & 1 \\
-1 & 0
\end{pmatrix}.
\]
We first choose the excitation ratio as $r_{\rm exc}=\frac 1 2$. Sampling a large number of eigenstates for different chain lengths $L$, we obtain the result as shown in Figure~\ref{fig:GEhalf}. It seems that the typical local distance $D$ between the eigenstate and the Gibbs ensemble becomes small in the thermodynamic limit $L\to\infty$ (potentially converging to zero).
Note that the maximum observed distance at each particle number decreases as the particle number is increased.

\begin{figure}[h]
	\centering
	\includegraphics[width=\linewidth]{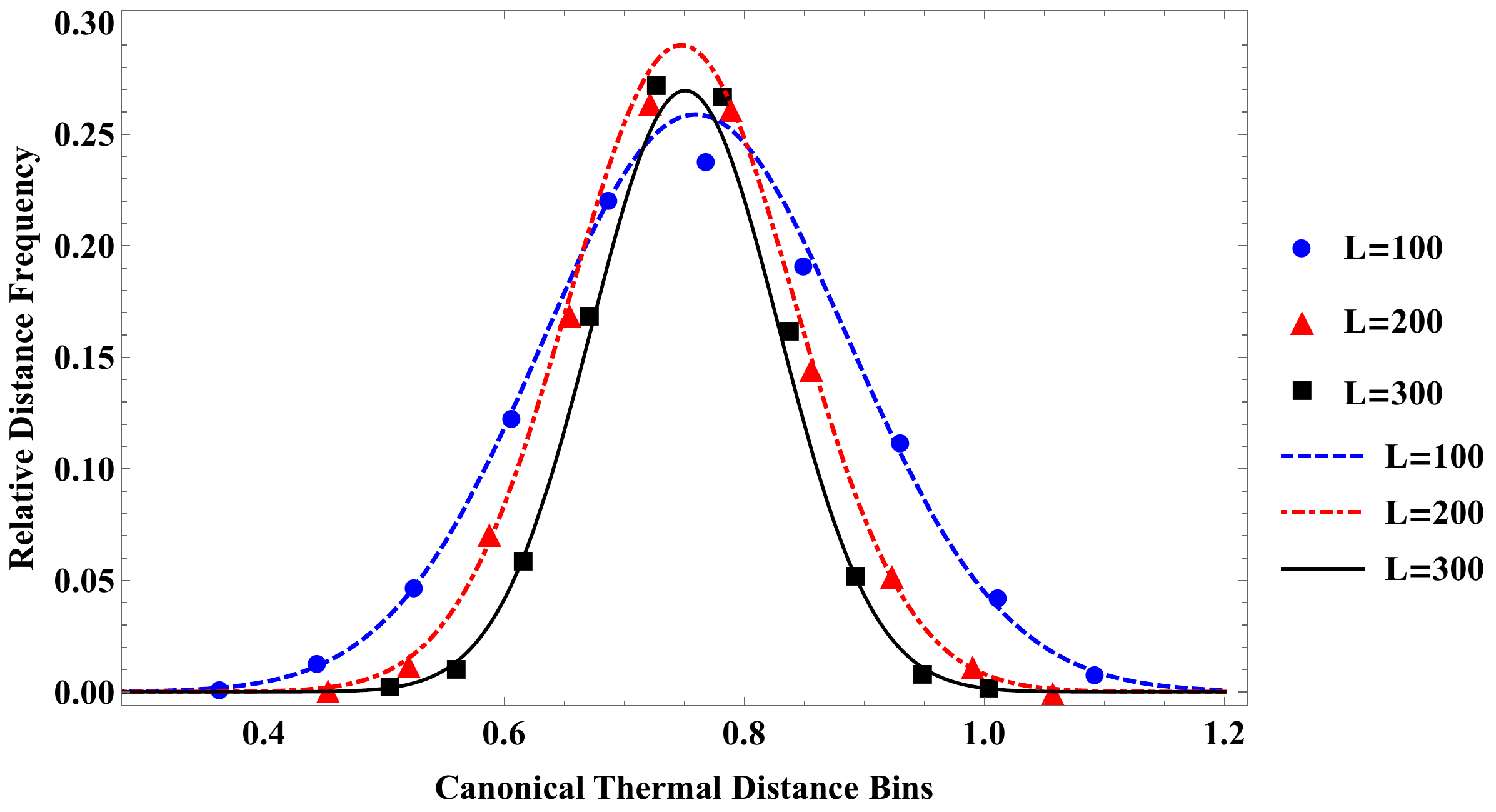}
	\caption{Frequency plot for Gibbs ensemble compared locally to sampled eigenstates with excitation ratio $r_{\rm exc}=\frac{1}{4}$. For each chain length $L$, 1800 eigenstates were sampled.}
	\label{fig:GEq}
\end{figure}

Before explaining this result with our analytic results of the previous sections, let us repeat the numerics with a different excitation ratio, namely $r_{\rm exc}=\frac 1 4$. The results are plotted in Figure~\ref{fig:GEq}.
The graph shows what appears to be the local distances converging to a non-zero value as the chain length is increased, pointing towards the eigenstates at this excitation ratio not converging to the Gibbs ensemble locally. Expanding on this result we can get a general idea how this works for all excitation ratios.

\begin{figure}[h]
	\centering
	\includegraphics[width=\linewidth]{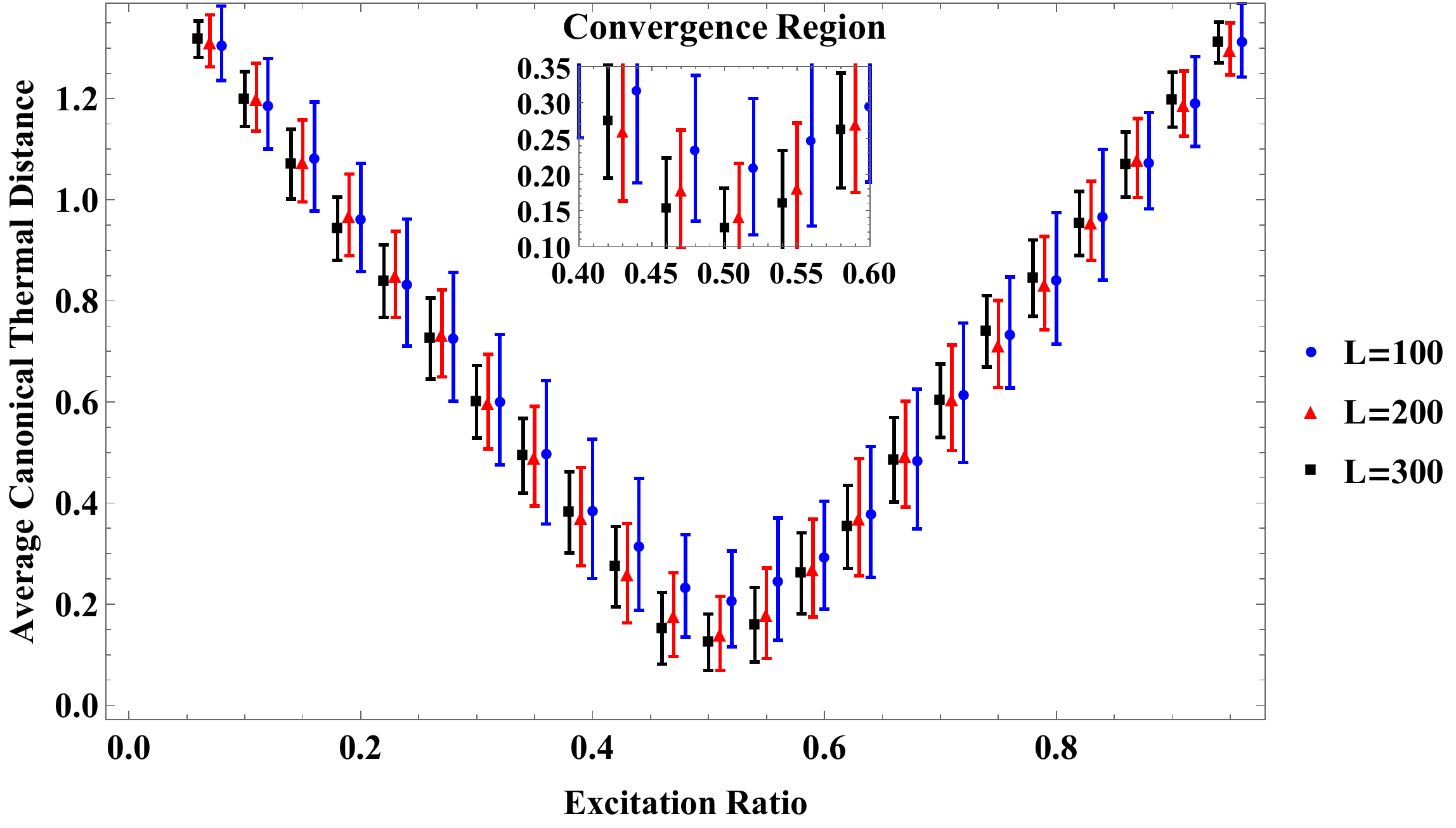}
	\caption{Average local distance $D$ between sampled eigenstates and Gibbs ensemble plotted against excitation ratio. We have sampled 100 eigenstates at each excitation ratio. Here and in all of the following plots, the error bars indicate the standard deviation of the distance $D$ as numerically determined from the samples. It turned out that these error bars do not visibly change if we increase the number of samples, which means that they faithfully represent the scattering of the values over all eigenstates of the corresponding property (here of all eigenstates with the given excitation ratio). Note that we have plotted $D$ for the same sets of excitations ratios for all lengths $L$, but we have horizontally shifted the $L=200$ and $L=300$ bars by a few pixels for better visibility.}
	\label{fig:GEer}
\end{figure}
Figure~\ref{fig:GEer} shows that for excitation ratios of about $r_{\rm exc}\approx \frac 1 2$, the numerics is consistent with local convergence between typical eigenstates and the Gibbs ensemble. For other excitation ratios, the average distance between the two does not appear to converge to zero. We thus observe another instance of the failure of eigenstate thermalization in integrable models, as pointed out before~\cite{Cassidy}.

So how can we understand this result analytically? Drawing eigenstates at random under a fixed $r_{\rm exc}$ corresponds to generating random eigenstates under a constraint of fixed particle number, $\langle\hat N\rangle=L\cdot r_{\rm exc}$. Thus, instead of the canonical ensemble, we will in general have to consider an ensemble that has $\hat N$ as one of its conserved quantities; we will do so in the next subsection. When we do not take the conserved quantity $\langle \hat N\rangle$ into account in the construction of the GGE, there is no reason to expect that the eigenstates will locally resemble that GGE (which is in this case just the GE).

However, the case of $r_{\rm exc}=\frac 1 2$ is special: it is the excitation ratio of \emph{unconstrained} typical states. This can be seen as follows. Suppose we draw an energy eigenstates uniformly at random from \emph{all} $2^L$ eigenstates of $\hat H$, without any restriction. Then the resulting state should locally reproduce the predictions of the maximally mixed or \emph{infinite temperature} state $\rho_{\beta=0}=2^{-L}\mathbf{1}$, since this is the GGE if there are no conserved quantities at all. But since $\tr(\hat d_k^\dagger d_k)=2^{L-1}$, the expected particle number in that state is
\[
   \langle \hat N\rangle=\tr(\hat N \rho_{\beta=0})=2^{-L}\tr\hat N = 2^{-L}\cdot L \cdot 2^{L-1}=L/2.
\]
Thus, fixing the particle number to $\langle \hat N\rangle:= L/2$ will statistically, on average for many samples, have the same effect as not fixing any constraint at all. Therefore, in this special case, it is correct not to invoke $\hat N$ as a conserved quantity in the construction of the corresponding GGE. We can take the GE to approximate local expectation values, or we can take the maximally mixed state which is the GGE for the case that there are no conserved quantities at all.

\subsection{Grandcanonical ensemble}
The next ensemble we investigate is the grandcanonical ensemble, which conserves energy and particle number,  $\hat H$ and $\hat N$. The mode covariance matrix for the grandcanonical ensemble is
\[
\Lambda_{GCE}^d = \bigoplus_{k=1}^L \tanh\left(\frac{\beta_1\epsilon_k}{2}+\frac{\beta_2}{2}\right)
\begin{pmatrix}
0 & 1 \\
-1 & 0
\end{pmatrix}.
\]
We proceed by repeating the same numerical experiments as the previous subsection. 

\begin{figure}[h]
	\centering
	\includegraphics[width=\linewidth]{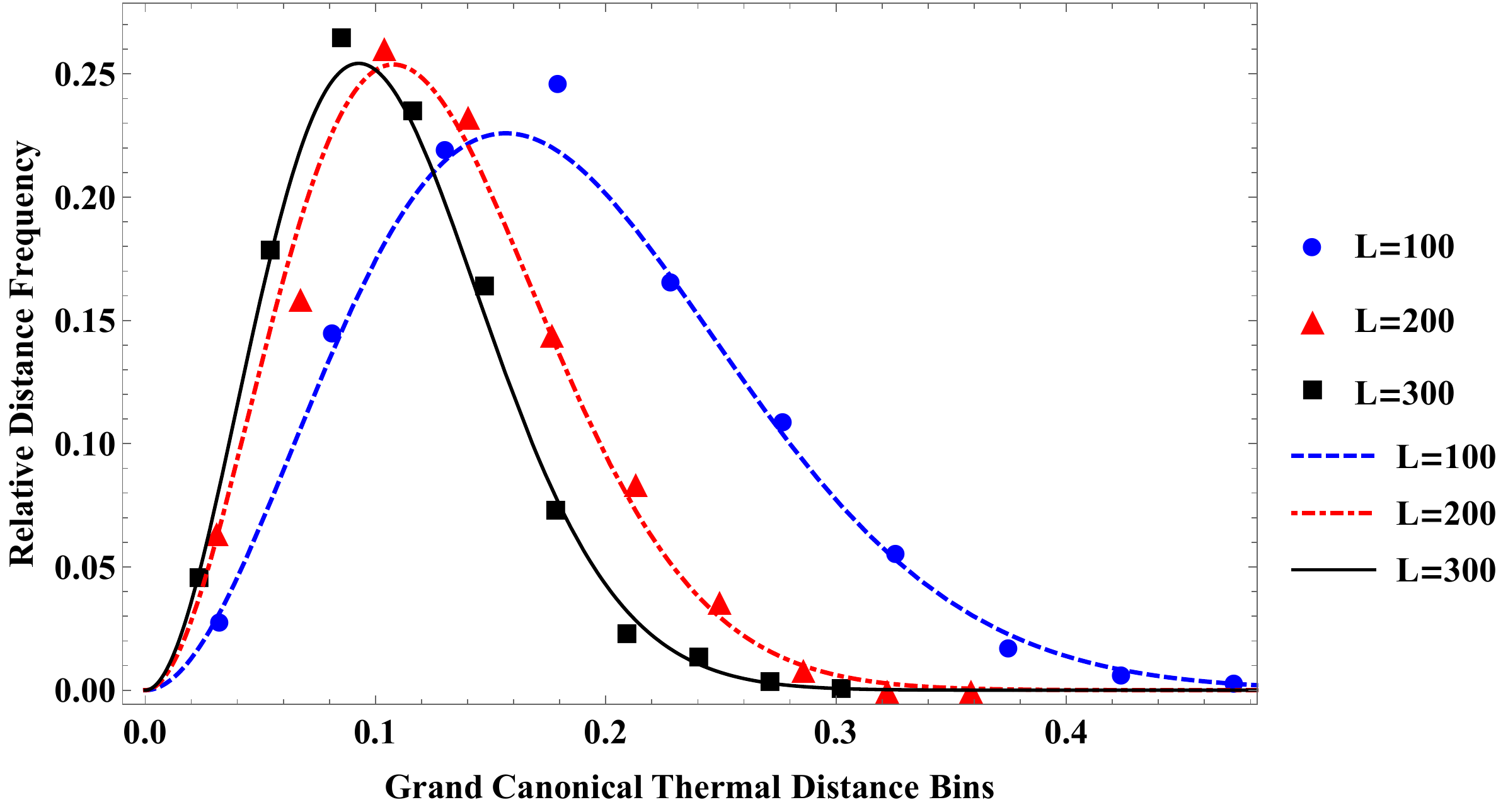}
	\caption{Distance frequency plot for grandcanonical ensemble compared locally to sampled eigenstates with excitation ratio $r_{\rm exc}=\frac{1}{2}$. For each chain length $L$, 1800 eigenstates were sampled.}
	\label{fig:GChalf}
\end{figure}

\begin{figure}[h]
	\centering
	\includegraphics[width=\linewidth]{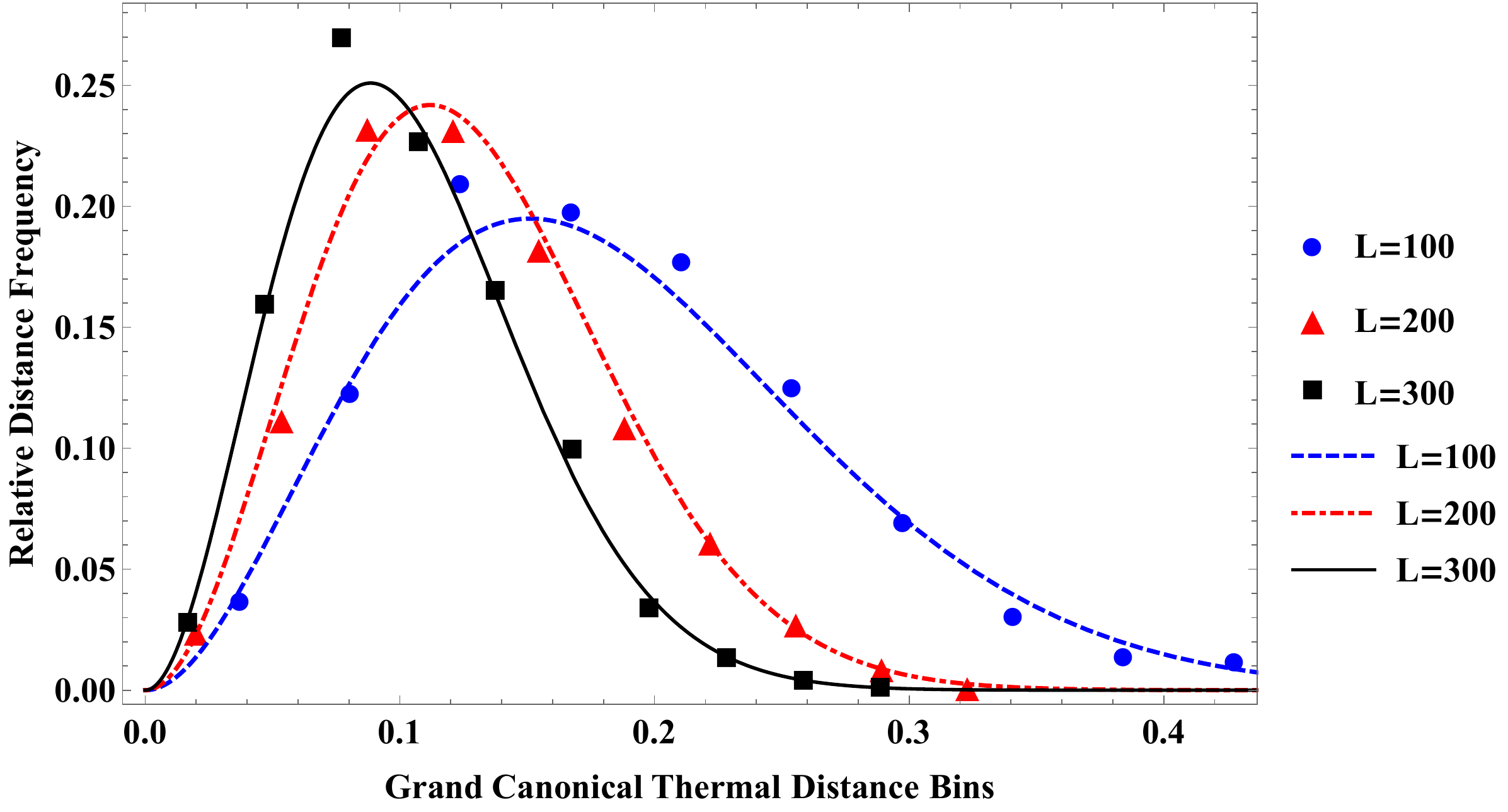}
	\caption{Distance frequency plot for grandcanonical ensemble compared locally to sampled eigenstates with excitation ratio $r_{\rm exc}=\frac{1}{4}$.  For each chain length $L$, 1800 eigenstates were sampled. }
	\label{fig:GCq}
\end{figure}

\begin{figure}[h]
	\centering
	\includegraphics[width=\linewidth]{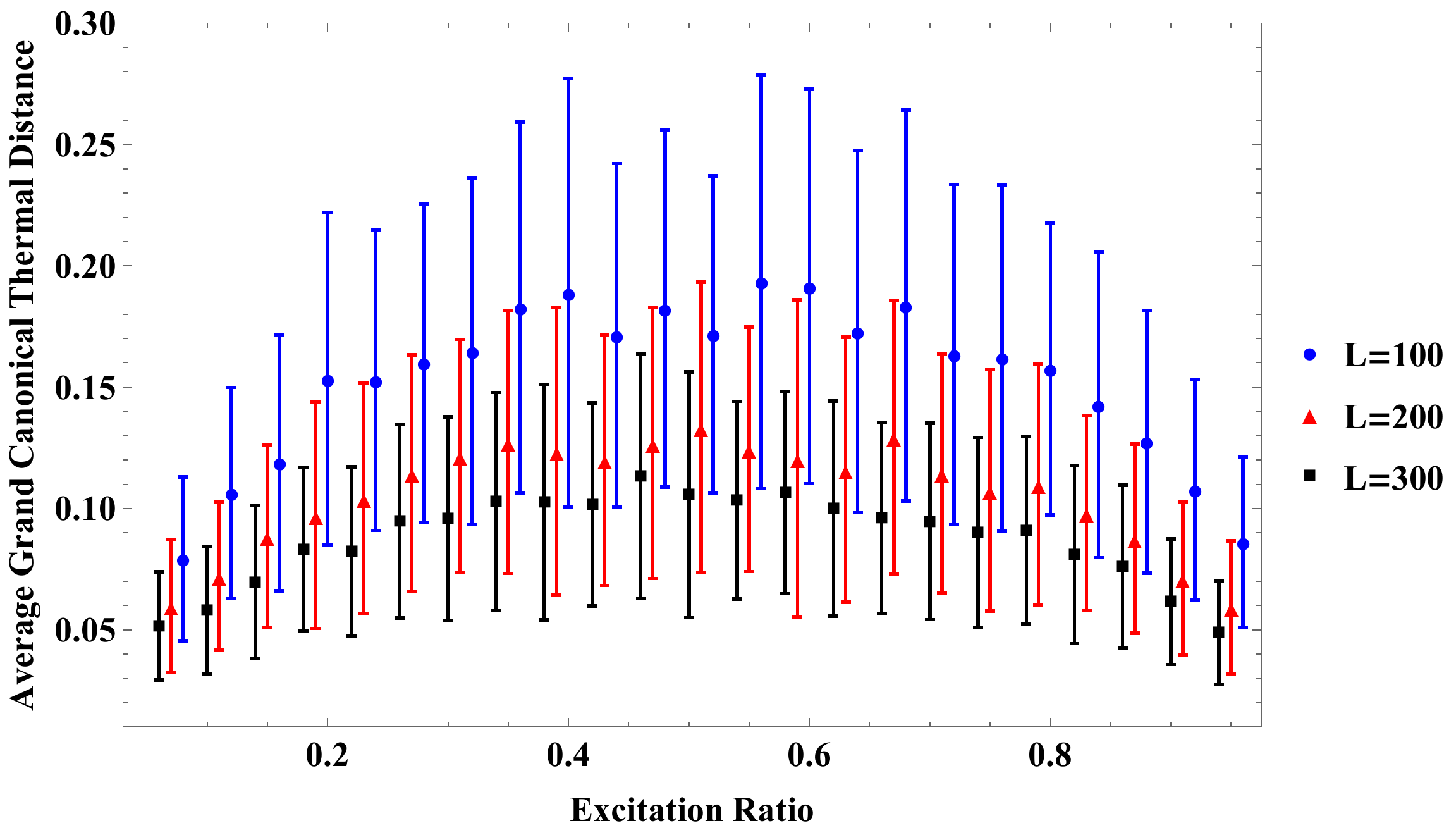}
	\caption{Average local distance plot between sampled eigenstates and grandcanonical ensemble plotted against excitation ratio. 100 sampled eigenstates at each excitation ratio. }
	\label{fig:GCer}
\end{figure}

Figures~\ref{fig:GChalf}, \ref{fig:GCq} and \ref{fig:GCer} show that the grandcanonical ensemble is the appropriate ensemble to use for the current sampling method at all excitation ratios. We observe that on average eigenstates get closer locally to the grandcanonical ensemble with growing $L$. This numerical test confirms the analytic results of Section~\ref{SecGET} (which have already been shown by Lai and Yang~\cite{Lai} in the special case of the grandcanonical ensemble): here we draw states at random under fixed particle number $\hat N$; by postselecting on their final energy, we can also consider the energy $\hat H$ to be fixed in retrospect, and then the grandcanonical ensemble will correctly describe the local statistics of typical eigenstates.

Nevertheless, if we draw eigenstates according to even more conserved quantities, the grandcanonical ensemble will lose its relevance, and we have to go beyond the results of Lai and Yang~\cite{Lai}. We will now generate eigenstates by picking an excitation ratio, and then randomly distributing this ratio according to predefined frequencies on the left and right half of the list of possible excitations. The previous experiments would have approximately half of the excitations on the left side of the list of excitations. (A more formal description will follow in Subsection~\ref{SubsecNumericalGGE} below.)
\begin{figure}[h]
	\centering
	\includegraphics[width=\linewidth]{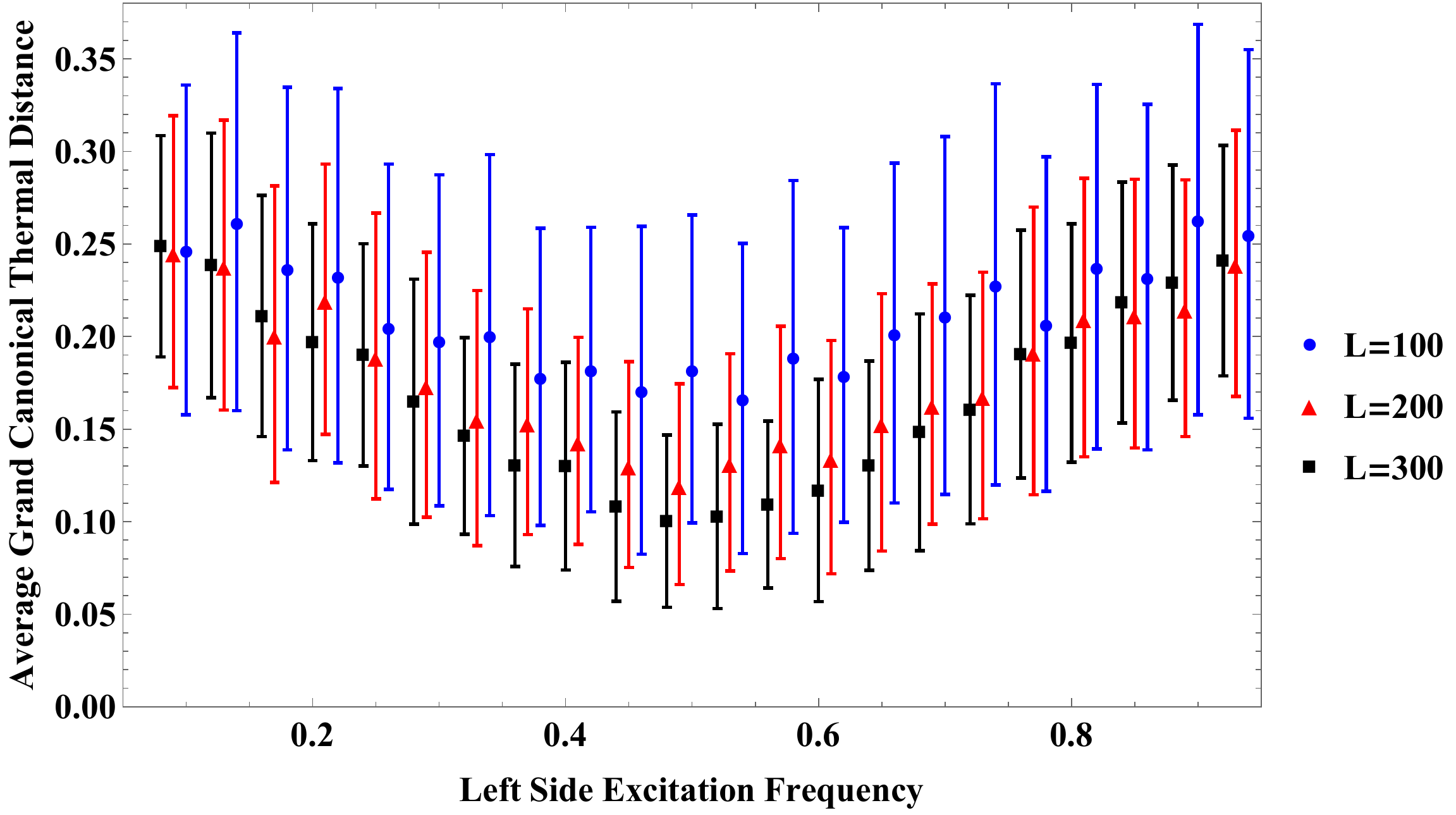}
	\caption{Average local distance between sampled eigenstates with $r_{\rm exc}=\frac{2}{5}$ and grandcanonical ensemble, plotted against left side frequency of excitation distribution. 100 eigenstates sampled at each test left side frequency.}
	\label{fig:GCls}
\end{figure}

In Figure~\ref{fig:GCls}, we see the ensemble tested against eigenstates with excitation ratio $r_{\rm exc} = \frac{2}{5}$ sampled at different left side excitation frequencies. As expected in the middle of the graph, where this experiment is equivalent to the previous sampling methods, we see on average convergence. However, moving away from this region, we see that the ensemble becomes worse at predicting the local statistics of our eigenstates. We can observe this phenomenon more closely by creating a distance frequency plot with a fixed excitation ratio and left side frequency. 

\begin{figure}[h]
	\centering
	\includegraphics[width=\linewidth]{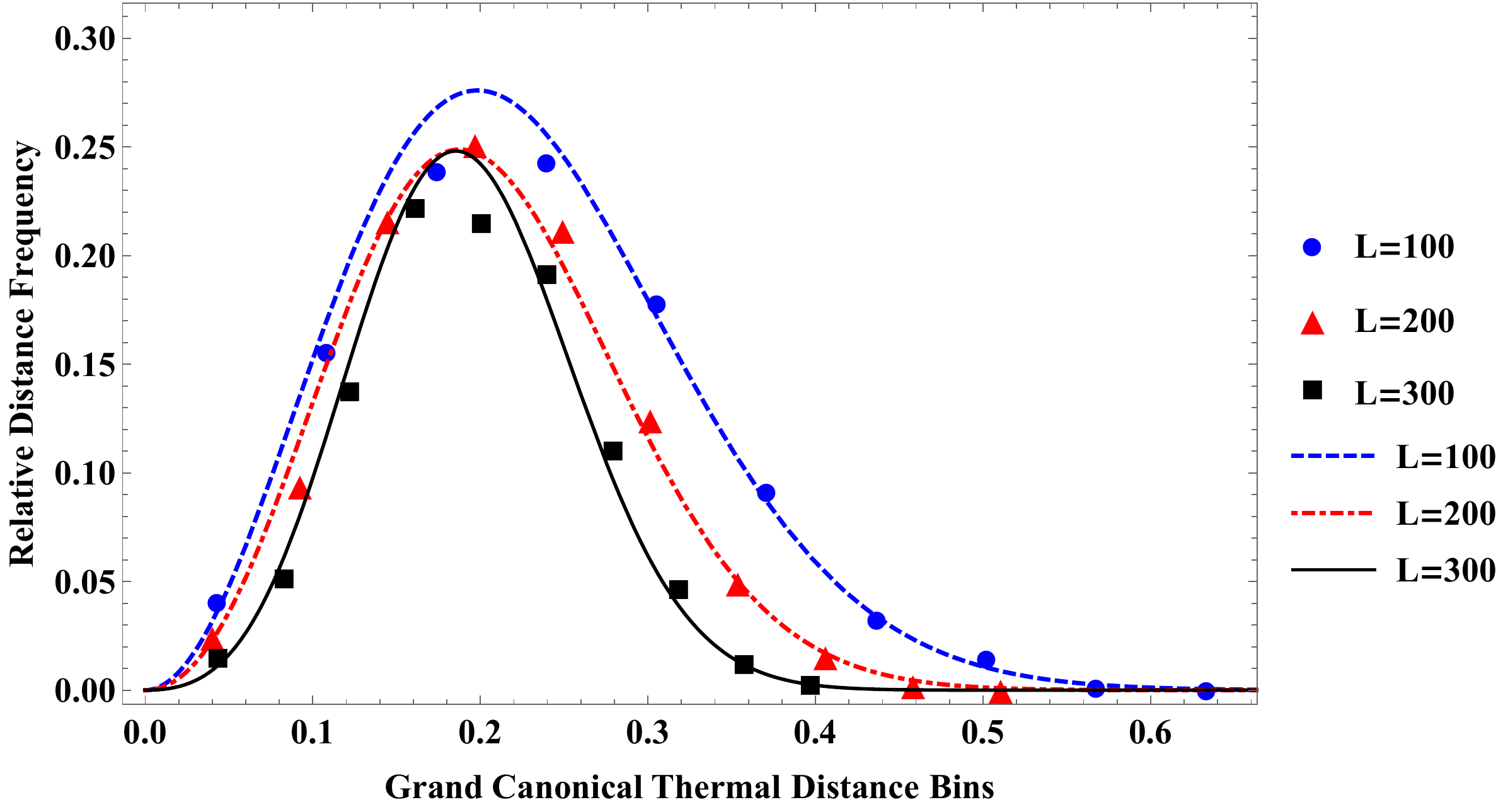}
	\caption{Distance frequency plot for grandcanonical ensemble compared locally to sampled eigenstates with excitation ratio $r_{\rm exc}=\frac{2}{5}$ and left side excitation frequency $\frac{4}{5}$. For each chain length $L$, 1800 eigenstates were sampled.}
	\label{fig:GC25}
\end{figure}
Similar to Figure~\ref{fig:GEq}, Figure~\ref{fig:GC25} appears to show that the eigenstates converge to a non-zero distance away from the ensemble. So with this sampling technique we have produced eigenstates that do not locally resemble the grandcanonical ensemble. As we will now see, the results of Section~\ref{SecGET} allow to describe the resulting ensemble of eigenstates in terms of a further generalization of the grandcanonical ensemble.

Again we wish to expand the ensemble to encompass the eigenstates of this new sampling method. The analytical results of this paper allow the coefficients of the conserved quantities to appear piecewise continuous in the thermodynamic limit, a fact we can take advantage of by splitting the number operator into two operators.

\subsection{Generalized Gibbs ensemble}
\label{SubsecNumericalGGE}
For our final numerical subsection, we introduce a generalized Gibbs ensemble that corrects the failures of the grandcanonical ensemble observed in Figures~\ref{fig:GCls} and \ref{fig:GC25}. Restricting ourselves to an even number of sites $L$, we consider the conserved quantities $\hat{H}=\sum_k\epsilon_k(\hat{d}_k^\dagger \hat{d}_k-\frac{1}{2})$, $\hat{N}_1=\sum_k q_{1,k}(\hat{d}_k^\dagger \hat{d}_k-\frac{1}{2})$ and $\hat{N}_2=\sum_k q_{2,k}(\hat{d}_k^\dagger \hat{d}_k-\frac{1}{2})$, where
\[
q_{1,k} = \left\{
\begin{array}{cl}
1 & \text{for }  k\leq \frac{L}{2}  \\
0 &\text{for }   k > \frac{L}{2} 
\end{array}\right.,\qquad
q_{2,k} = \left\{
\begin{array}{cl}
0 & \text{for }  k\leq \frac{L}{2}  \\
1 &\text{for }   k > \frac{L}{2} 
\end{array}\right. ,
\]
with the eigenstates labelled such that $\epsilon_k\leq\epsilon_{k+1}$. The covariance matrix of the GGE becomes
\[
\Lambda_{GGE}^d = \bigoplus_{k=1}^n \tanh\left(\frac{\beta_1\epsilon_k}{2}+\frac{\beta_2 q_{1,k}}{2}+\frac{\beta_3 q_{2,k}}{2}\right)
\begin{pmatrix}
0 & 1 \\
-1 & 0
\end{pmatrix}.
\]
\begin{figure}[h]
	\centering
	\includegraphics[width=\linewidth]{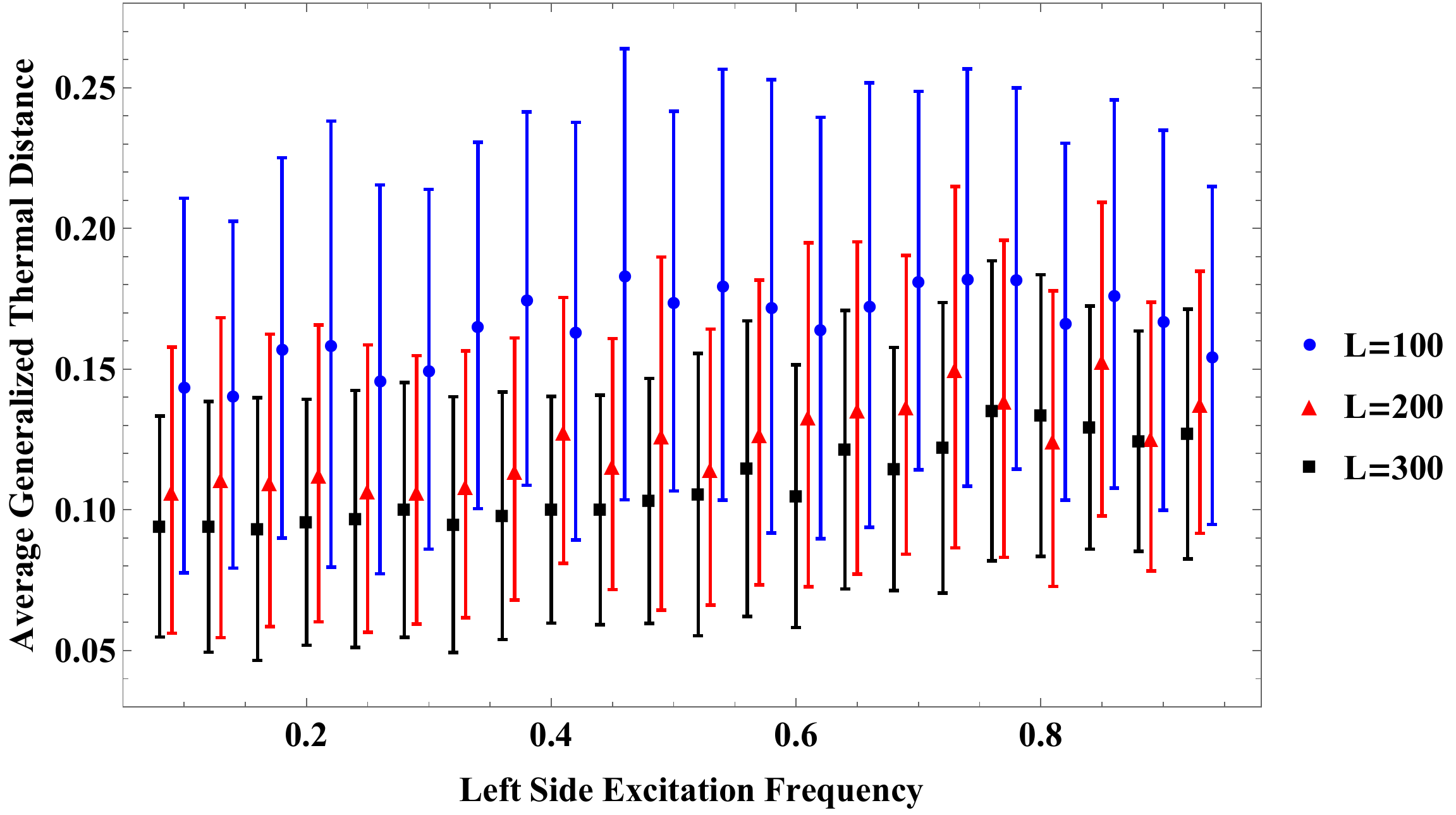}
	\caption{Average local distance plot between sampled eigenstates with $r_{\rm exc}=\frac{2}{5}$ and generalized Gibbs ensemble plotted against left side frequency of excitation distribution. 100 eigenstates sampled at each test left side frequency.}
	\label{fig:GGls}
\end{figure}

\begin{figure}[h]
	\centering
	\includegraphics[width=\linewidth]{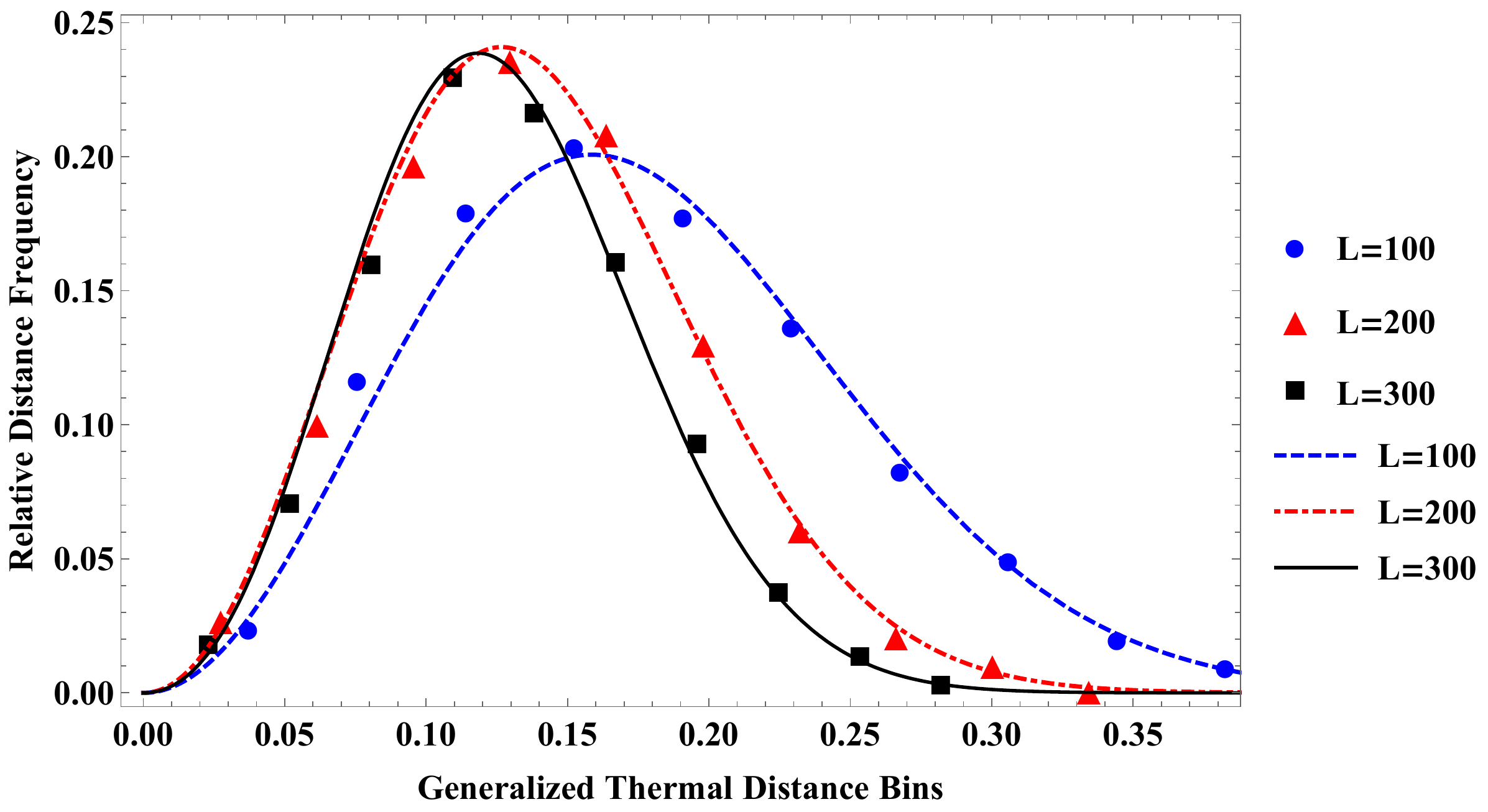}
	\caption{Distance frequency plot for generalized Gibbs ensemble compared locally to sampled eigenstates with excitation ratio $r_{\rm exc}=\frac{2}{5}$ and left side frequency $\frac{4}{5}$. 1800 eigenstates were sampled at each chain length $L$.}
	\label{fig:GG25}
\end{figure}
Now we sample eigenstates with fixed excitation ratio $r_{\rm exc}$ and ratio of excitations in the ``left bin'' $r_{\rm left}$. That is, every eigenstate is generated by applying $L\cdot r_{\rm exc}$ random creation operators $\hat d_k^\dagger$ to the vacuum, but exactly $L\cdot r_{\rm exc}\cdot r_{\rm left}$ of those excitations are chosen such that $0\leq k \leq L/2$. This is equivalent to fixing the values of $\hat N_1$ and $\hat N_2$, and so our analytic result of Section~\ref{SecGET} claims that the resulting states will typically be locally close to the corresponding GGE.

This is indeed what Figures~\ref{fig:GGls} and \ref{fig:GG25} show. In Figure~\ref{fig:GGls} we see that the average distances around the ``typical'' excitation distribution $r_{\rm left}\approx \frac 1 2$ agree with Figure~\ref{fig:GCls}, but away from this typical value the generalized Gibbs ensemble performs much better than the grandcanonical ensemble. Figure~\ref{fig:GG25} also shows signs of average convergence. Thus, our numerics confirm the analytic findings of this paper: to accurately describe random eigenstates locally in terms of some statistical ensemble, one has to build the GGE corresponding to the quantities that have been held fixed in the sampling process.

\section{Conclusions}
We have analytically shown that quasifree fermionic models satisfy a weak generalized version of the ETH: the vast majority of eigenstates which arise from unbiased sampling according to a finite number of constraints are locally close to the corresponding GGE. The conserved quantities held fixed are assumed to be of the form~(\ref{eqConstraints}), which includes the total energy, particle number, as well as other quantities like the one we have considered in Subsection~\ref{SubsecNumericalGGE}. We have also illustrated our results numerically by example of the XX spin chain, which can be written as a fermionic model by means of a Jordan-Wigner transformation.

Our results give further evidence to the hypothesis that the GGE is the correct ensemble to describe the emergence of thermalization in integrable models. Previous work has focused on constructing the GGE from the full set of all conserved quantities. However, in the case of quasifree fermionic models, this includes \emph{all} mode excitations $\hat d_{\vec k}^\dagger \hat d_{\vec k}$ --- the number of these operators grows extensively with system size. What we have shown is that the GGE attains its relevance already in the simpler situation that a small and finite number of conserved quantities is fixed. In this case, the vast majority of energy eigenstates satisfying these constraints is locally well described by the corresponding GGE, constructed from maximizing the entropy with respect to this finite number of constraints.

\section*{Acknowledgments}
We are grateful to Llu\'is Masanes and Henrik Wilming for discussions, and to Alex Buchel and David Jeffrey for help with organizational aspects of this project. We thank the Natural Science \& Engineering Research Council of Canada for financial support. This research was undertaken, in part, thanks to funding from the Canada Research Chairs program. Also, this research was supported in part by Perimeter Institute for Theoretical Physics. Research at Perimeter Institute is supported by the Government of Canada through the Department of Innovation, Science and Economic Development Canada and by the Province of Ontario through the Ministry of Research, Innovation and Science.

\onecolumngrid

\section*{Appendix}
\subsection{Generalized Gibbs covariance matrix}
\label{SubsecGGE}
Generalizing the derivation of~\cite{Wilming}, we find an expression for the covariance matrix of the generalized Gibbs ensemble. Suppose we start with  a generalized Gibbs ensemble of the form
\[
	\rho_{\textsc{GGE}} = \frac{e^{-\sum_{i}\beta_i \hat Q_i'}}{Z},
\]
where $\hat Q'_i=\sum_{\vec{k}} 2 i q'_{i,\vec{k}} \check d_{\vec{k},1}\check d_{\vec{k},2}$. Setting $A_{\vec{k}}:=2 i \check d_{\vec{k},1}\check d_{\vec{k},2}$, we have $A_{\vec{k}}^2=\mathbf{1}$, and thus for all $\alpha\in\R$
\[
	 e^{\alpha A_{\vec k}} = \sum_{j=0}^{\infty}\left(\frac{(\alpha A_{\vec k})^{2j}}{(2j)!}+\frac{(\alpha A_{\vec k})^{2j-1}}{(2j-1)!}\right)=(\cosh \alpha)\mathbf{1}+(\sinh \alpha)A_{\vec k}.
\]
Note that $[A_{\vec k},A_{\vec l}]=0$ for all $\vec{k},\vec{l}$, hence $[\hat Q'_i,\hat Q'_j]=0$ for all $i,j$. This allows us to write
\begin{equation}
   e^{-\sum_i \beta_i \hat Q'_i}=\prod_{\vec k} e^{-\sum_i \beta_i q'_{i,\vec{k}} A_{\vec k}}
   =\prod_{\vec k}\left[ \left(\cosh\sum_i \beta_i q'_{i,\vec{k}}\right)\mathbf{1}-\left(\sinh \sum_i \beta_i q'_{i,\vec{k}}\right)A_{\vec k}\right].
   \label{eqComplicated}
\end{equation}
Next we must calculate the form of the partition function $Z$. This needs some preparation. First, suppose that $\vec{k}_1,\ldots,\vec{k}_n$ are $n$ pairwise distinct momentum vectors, i.e.\ $\vec{k}_i\neq \vec{k}_j$, then
\begin{equation}
	\tr(A_{\vec{k}_1}A_{\vec{k}_2}\ldots A_{\vec{k}_n})=0.
	\label{eqPrep}
\end{equation}
To prove this, note that the left-hand side is, up to a constant factor, equal to the following expression, to which we apply first the cyclicity of the trace and then the anticommutation relations of the Majorana operators:
\[
	\tr(\check d_{\vec{k}_1,1}\check d_{\vec{k}_1,2}\check d_{\vec{k}_2,1}\check d_{\vec{k}_2,2}\ldots \check d_{\vec{k}_n,1}\check d_{\vec{k}_n,2})=
		\tr(\check d_{\vec{k}_1,2}\check d_{\vec{k}_2,1}\check d_{\vec{k}_2,2}\ldots \check d_{\vec{k}_n,1}\check d_{\vec{k}_n,2} \check d_{\vec{k}_1,1})
		=-\tr(\check d_{\vec{k}_1,2}\check d_{\vec{k}_2,1}\check d_{\vec{k}_2,2}\ldots \check d_{\vec{k}_n,1} \check d_{\vec{k}_1,1} \check d_{\vec{k}_n,2}).
\]
We go on by anticommuting the term $\check d_{\vec{k}_1,1}$ further to the left, with every step yielding a minus sign. In the end, we will reproduce the original expression, but with an extra overall minus sign. This proves~(\ref{eqPrep}).

By multiplying out the right-hand side of±(\ref{eqComplicated}) (using again that the $A_{\vec k}$ commute pairwise), we obtain an expression of the form
\[
   e^{-\sum_i \beta_i \hat Q'_i}=\prod_{\vec k}\cosh\left(\sum_i \beta_i q'_{i,\vec{k}}\right)\mathbf{1}+\sum_n \sum_{\vec{k}_1,\ldots,\vec{k}_n} c_{\vec{k}_1,\ldots,\vec{k}_n} A_{\vec{k}_1}\ldots A_{\vec{k}_n},
\]
where the $\vec{k}_1,\ldots,\vec{k}_n$ on the right-hand side are pairwise distinct. Thus, taking the trace and using~(\ref{eqPrep}), we obtain
\[
   Z=\tr\left( e^{-\sum_i \beta_i \hat Q'_i}\right)=2^{L^d} \prod_{\vec k}\cosh\left(\sum_i \beta_i q'_{i,\vec{k}}\right).
\]
This gives the generalized Gibbs ensemble
\[
   \rho_{\rm GGE}=\prod_{\vec k}\left[ \frac 1 2 \mathbf{1}-\frac 1 2 \left(\tanh \sum_i \beta_i q'_{i,\vec{k}}\right)A_{\vec k}\right].
\]

With this form we can investigate the covariances:

\[
   \Lambda^d_{(\vec{k},a),(\vec{k'},b)}=2i\,\tr \left(\check d_{\vec{k},a}\check d_{\vec{k'},b} 2^{-L^d}\left(\mathbf{1}+\sum_{m=1}^{L^d}\sum_{\vec{l}_1,\ldots,\vec{l}_m} c_{\vec{l}_1,\ldots,\vec{l}_m} A_{\vec{l}_1} A_{\vec{l}_2}\ldots A_{\vec{l}_m}\right)\right),
\]
where $c_{\vec{l}_1,\ldots,\vec{l}_m}\in\mathbb{R}$ are constants. To deduce the expression of the covariances we must take a few more steps and note a few more relationships. First, note that $\check d_{\vec{k},a}^2 = \frac{1}{2}$.  Recalling that $A_{\vec{k}}^2=\mathbf{1}$ and $A_{\vec{k}}=2 i \check d_{\vec{k},1}\check d_{\vec{k},2}$ as well as~(\ref{eqPrep}), we conclude that
	\[\Lambda_{(\vec{k},a),(\vec{k},a)}^d = 0 \quad\text{ and }\quad \Lambda_{(\vec{k},2),(\vec{k},1)}^d = -\Lambda_{(\vec{k},1),(\vec{k},2)}^d = \tanh\left(\sum_i \beta_i q'_{i,\vec{k}}\right) .
	\]
It remains to show that $\Lambda^d_{(\vec{k},a),(\vec{k}',b)}=0$ for $\vec{k}\neq\vec{k}'$. So we must prove the following:
\begin{equation}
\tr(\check d_{\vec{k},a}\check d_{\vec{k'},b}A_{\vec{l}_1}A_{\vec{l}_2}\ldots A_{\vec{l}_n})=0\qquad\mbox{if }\vec{k}\neq\vec{k}'.
\label{eqPrep1}
\end{equation}
Two cases have to be distinguished, depending on the composition of the product of momentum vectors. First suppose that the momentum vectors $\vec{k}$ and $\vec{k'}$ are unique to the list, that is $\vec{k}\neq\vec{l}_i$ and $\vec{k}'\neq\vec{l}_i$ for all $i$. Then we have an even number of pairwise different Majorana operators inside the trace, and~(\ref{eqPrep1}) follows from the anticommutation relations with a similar calculation as the proof of~(\ref{eqPrep}). Second, suppose that there is some $i$ such that $\vec{k}=\vec{l}_i$. Then we can use the anticommutation relations and cyclicity of the trace to move the $\check{d}_{\vec{k},a}$ next to the $A_{\vec{l}_i}=A_{\vec{k}}$ and then apply $\check{d}_{\vec{k},a}^2=\frac 1 2$ to effectively get rid of the $\check{d}_{\vec{k},a}$ and the $\check{d}_{\vec{l}_i,a}$. If there is some $j$ such that $\vec{k}'=\vec{l}_j$ we do the same. Thus, we recover the case of an even number of pairwise distinct Majorana operators in the trace, and~(\ref{eqPrep1}) follows. Since $q'_{i,\vec{k}}=-\frac 1 2 q_{i,\vec{k}}$, this proves the form of the covariance matrix as claimed in the main text.

\end{document}